\documentclass{jfm}
\usepackage{amsmath,amsfonts,amssymb,amsgen,amsbsy,upmath,bm}
\newcommand{\ud}{\mathrm{d}}
\newcommand{\ue}{\mathrm{e}}
\newcommand{\ui}{\mathrm{i}}

\title[Point-source inertial particle dispersion]{Point-source inertial particle dispersion}
\author[M.\ Martins Afonso and A.\ Mazzino]{M\ls A\ls R\ls C\ls O\ns M\ls A\ls R\ls T\ls I\ls N\ls S\ns A\ls F\ls O\ls N\ls S\ls O$^1$\ns \and A\ls N\ls D\ls R\ls E\ls A\ns M\ls A\ls Z\ls Z\ls I\ls N\ls O$^2$}
\affiliation{$^1$Institut~de~M\'ecanique~des~Fluides~de~Toulouse~-~groupe~Ecoulements~Et~Combustion, all\'ee~du~Professeur~Camille~Soula, 31400~Toulouse, France\\[\affilskip]$^2$Department~of~Physics~-~University~of~Genova, and CNISM \& INFN~-~Genova~Section, via~Dodecaneso~33, 16146~Genova, Italy}
\date{\today}

\begin{document}

 \maketitle

 \begin{abstract}
  The dispersion of inertial particles continuously emitted from a point source is analytically investigated in the limit of small inertia. Our focus is on the evolution equation of the particle joint probability density function $p(\bm{x},\bm{v},t)$, $\bm{x}$ and $\bm{v}$ being the particle position and velocity, respectively. For finite inertia, position and velocity variables are coupled, with the result that $p(\bm{x},\bm{v},t)$ can be determined by solving a partial differential equation in a $2d$-dimensional space, $d$ being the physical-space dimensionality. For small inertia, $(\bm{x},\bm{v})$-variables decouple and the determination of $p(\bm{x},\bm{v},t)$ is reduced to solve a system of two standard forced advection--diffusion equations in the space variables $\bm{x}$. The latter equations are derived here from first principles, i.e.\ from the well-known Lagrangian evolution equations for position and particle velocity.
 \end{abstract}

 \section{Introduction}

 The study of particulate matter (PM) in flowing fluids is a problem attracting much attention because of its myriad of applications in different realms of science and technology. In way of example, we mention the implications on: climate dynamics and hydrological cycles (\cite{IPCC07}; \cite{LF05}), mainly in connection to global climate changes originated by PM-induced cloud formation (\cite{CMT08}); environmental sciences (\cite{SEKH01}), in relation to pollution and deterioration of visibility; epidemiology (\cite{D93}), in connection to adverse health effects in humans; and, finally, classical fluid dynamics, e.g.\ to understand how a flow field influences the particle concentration (\cite{EF94}; \cite{BFF01}; \cite{WM03}; \cite{B05}; \cite{BCH07}).

 The complex character of this discipline simply arises from the many interactions between totally different fields of science and technology. If, on the one hand, key achievements have been obtained on both the epidemiological aspect and on the side of experimental studies, aiming at eliciting information on the distribution and number concentration in the environment, on the other hand very little is known about the dynamical aspect of particulate matter in situations of interest. As an instance, the basic equations ruling the spatio-temporal evolution of the particle concentration field are up to now unknown in the presence of paradigmatic environmental sources, such as, e.g., point- and line-source emissions (mimicking the release from a chimney and from a street, respectively). This seems to be the case in spite of the fact that the basic research dealing with inertial particles in fluids is a field in strong development, and where important results have been obtained in the last few years (see, e.g., \cite{WM03}; \cite{RMP04}; \cite{PS05}).

 Our specific aim here is to make a first step in the intermediate framework lying between the realm of abstract models of transport of inertial particles (sometimes too abstract to have direct implications on applicative fields) and that of empirical models of transport (sometimes too empirical and with scarce contact with the underlying physics). Along this way, we will focus on the dynamics of PM when emitted from a point source. Note that only recently the point-source emission for the case of neutral particles (i.e.\ having the same density as the surrounding fluid) has been addressed quantitatively from first principles (\cite{CMAM07}). The main difference between the cases of neutral and of inertial particles is that, in the former case, the equations for the space/time evolution of particle concentration are very well known, and the resulting phenomenology has been elucidated only recently; in the latter case, we are still waiting for the governing field equations. Here, our goal is to deduce such equations from first principles, i.e.\ from the known Lagrangian evolution of inertial particles (\cite{MR83}).

 The paper is organized as follows. In \S~\ref{sec:pos} we sketch the problem under consideration by recalling the significant equations and by performing quantitative balances to justify our approach. In \S~\ref{sec:exp} we analyse our small-inertia expansion order by order, focusing on the related equations which stem out as solvability conditions.
 Conclusions follow in \S~\ref{sec:conc}. The appendix is devoted to briefly recalling the main steps leading to the Fokker--Planck equation for the phase-space density evolution, starting from the well-known Lagrangian equations ruling the particle evolution.

 \section{Position of the problem} \label{sec:pos}

 Let us consider the problem of dispersion of single, small, spherical inertial particles emitted from a localized source, such as a chimney releasing some pollutant in the atmosphere or as an injection point (a syringe) in a microchannel, at a rate $T^{-1}$. For the sake of simplicity, the spatial structure of the source is approximated as punctual, and located in the origin of our frame of reference. In terms of the release velocity, the emission distribution (denoted by $f$) will be left as unspecified (with the only constraint of normalization) for most of the general calculations, and then for exemplificative applications will be modelled as a Gaussian, centered on an average value $\bm{v}_*$ and with standard deviation $\sigma$. The axes are chosen such as to have $\bm{v}_*$ aligned with the positive $x_d$ direction (our polar axis): for the specific case of atmospheric pollution released from a smokestack, it vertically points upwards, i.e.\ opposed to the gravity acceleration $\bm{g}$, due to buoyancy effects. We consider as free parameters both the space dimension (even if the three-dimensional case should be born in mind for applicative purposes) and the coefficient $\beta$, built from the particle ($\rho_{\mathrm{p}}$) and the fluid ($\rho_{\mathrm{f}}$) densities as follows: $\beta\equiv3\rho_{\mathrm{f}}/(\rho_{\mathrm{f}}+2\rho_{\mathrm{p}})$.\\
 Neglecting some corrective terms but taking into account the added-mass effect in a simplified way (\cite{MA08}), the dynamical equations ruling the evolution of the particle position ($\bm{X}(t)$) and covelocity ($\bm{V}(t)$) in an incompressible flow $\bm{u}(\bm{x},t)$ read (\cite{MR83}):
 \begin{equation} \label{dyn}
  \left\{\begin{array}{l}
   \dot{\bm{X}}(t)=\bm{V}(t)+\beta\bm{u}(\bm{X}(t),t)\\
   \dot{\bm{V}}(t)=\displaystyle-\frac{\bm{V}(t)-(1-\beta)\bm{u}[\bm{X}(t),t]}{\tau}+(1-\beta)\bm{g}+\frac{\sqrt{2\kappa}}{\tau}\bm{\eta}(t)\;.
  \end{array}\right.
 \end{equation}
 The Stokes time $\tau\equiv R^2/(3\nu\beta)$ expresses the typical response delay of the particle to flow variations, i.e.\ the relaxation time in Stokes' viscous drag ($R$ and $\nu$ being the particle radius and the fluid viscosity, respectively); $\bm{\eta}(t)$ is the standard white noise associated to the particle Brownian diffusivity $\kappa$. Notice that, when $\beta\neq0$, there is a discrepancy between particle's velocity, $\dot{\bm{X}}(t)$, and covelocity, $\bm{V}(t)\equiv\dot{\bm{X}}(t)-\beta\bm{u}(\bm{X}(t),t)$; for steady flows, this means that, unless $\rho_{\mathrm{p}}\gg\rho_{\mathrm{f}}$, the values of $\bm{v}_*$ and $\sigma^2$ should be interpreted as $\overline{\dot{\bm{X}}}|_{\mathrm{emission}}-\beta\bm{u}|_{\bm{x}=\bm{0}}$ and $\overline{\dot{X}^2}|_{\mathrm{emission}}+\beta^2u|_{\bm{x}=\bm{0}}^2$, respectively (here the bar represents the average over the velocity distribution of the particle at the emission point).

 If we consider the external (unperturbed) flow as having characteristic length scale and velocity $L$ and $U$, the forced Fokker--Planck equation for the particle phase-space density $p(\bm{x},\bm{v},t)$ reads:
 \begin{equation} \label{fp}
  \left\{\ud_t+\partial_{\mu}\left(v_{\mu}+\beta u_{\mu}\right)+\nabla_{\mu}\left[\frac{(1-\beta)u_{\mu}-v_{\mu}}{\tau} +(1-\beta)g_{\mu}\right]-\frac{\kappa}{\tau^2}\nabla^2\right\}p=\frac{\delta(\bm{x})f(\bm{v})}{T}\;,
 \end{equation}
 where $\ud_t\equiv\partial/\partial t$, $\partial_{\mu}\equiv\partial/\partial x_{\mu}$ and $\nabla_{\mu}\equiv\partial/\partial v_{\mu}$. The derivation of this equation from (\ref{dyn}) is a lengthy but straightforward task, and is briefly recalled in the appendix.

 After performing the nondimensionalization (\cite{MA08})
 $$\bm{x}\mapsto\frac{\bm{x}}{L}\qquad\bm{u}\mapsto\frac{\bm{u}}{U}\qquad t\mapsto\frac{t}{L/U}\qquad\bm{v}\mapsto\frac{\bm{v}}{\sqrt{2\kappa/\tau}}\;,$$
 we can introduce the adimensional numbers
 $$\textrm{Stokes: }\mathrm{St}\equiv\frac{\tau U}{L}\;,\qquad\textrm{Froude: }\mathrm{Fr}\equiv\frac{U}{\sqrt{gL}}\;,\qquad\textrm{P\'eclet: }\mathrm{Pe}\equiv\frac{LU}{\kappa}\;,$$
 and the vertical unit vector pointing downwards: $\bm{G}\equiv\bm{g}/g$.\\
 Note that the particle covelocity $\bm{v}$ has a nondimensionalization different from the fluid velocity $\bm{u}$. Let us then try to explain the meaning of this nondimensionalization with $\sqrt{2\kappa/\tau}$, in view of our investigation in the limit of small inertia. In the phase space, this can be seen as resulting from a competition between the small-$\tau$ limit, which would reduce the phase-space dimension by making $\bm{v}$ collapse onto the variety $\bm{u}(\bm{x},t)$, and the effect of diffusivity, which counteracts this process. In order to have the nondimensionalized covelocity $\sim O(1)$ (i.e., a meaningful nondimensionalization) and, at the same time, the dimensional one $\sim O(U)$ (because of the small-inertia limit, which is usually associated with small deviations of the particle from the underlying fluid trajectory), the following is a sufficient condition:
 \begin{equation} \label{loosely}
  \sqrt{\frac{2\kappa}{\tau}}\sim O(\bm{v})\sim U\ \Longrightarrow\ \sqrt{\frac{2}{\mathrm{St}\,\mathrm{Pe}}}\sim O(1)\ \Longrightarrow\ \textrm{(loosely)}\ \mathrm{St}\,\mathrm{Pe}\sim O(1)\;.
 \end{equation}

 \section{Expansion at small inertia} \label{sec:exp}

 In terms of the nondimensional variables introduced in the previous section, (\ref{fp}) becomes:
 \begin{eqnarray} \label{fpa}
  &\displaystyle\hspace{-2cm}\left\{-\mathrm{St}^{-1}\left(\nabla_{\mu}v_{\mu}+\frac{1}{2}\nabla^2\right)+\mathrm{St}^{-1/2}\left[\sqrt{\frac{2}{\mathrm{Pe}}}\,v_{\mu}\partial_{\mu}+\sqrt{\frac{\mathrm{Pe}}{2}}\,(1-\beta)u_{\mu}\nabla_{\mu}\right]\right.\hspace{1cm}\\
  &\displaystyle\hspace{1cm}\left.+\mathrm{St}^0\left[\ud_t+\beta u_{\mu}\partial_{\mu}\right]+\mathrm{St}^{1/2}\left[\sqrt{\frac{\mathrm{Pe}}{2}}\,\frac{1-\beta}{\mathrm{Fr}^2}G_{\mu}\nabla_{\mu}\right]\right\}p=\frac{L}{TU}\delta(\bm{x})f(\bm{v})\;.\hspace{-2cm}\nonumber
 \end{eqnarray}

 Let us now perform a small-Stokes expansion of the particle probability density function (PDF) and a Hermitianization of the problem in the spirit of \cite{MA08}:
 \begin{equation} \label{herm}
  p(\bm{x},\bm{v},t)=\sum_{n=0}^{\infty}\mathrm{St}^{n/2}p_n(\bm{x},\bm{v},t)\;,\qquad p_n(\bm{x},\bm{v},t)=\ue^{-v^2/2}\psi_n(\bm{x},\bm{v},t)\;.
 \end{equation}
 Notice that, because $p$ is normalized to unity, each $p_n$ must be normalized to $\delta_{n0}$. Plugging (\ref{herm}) into (\ref{fpa}), we obtain a chain of equations for $\psi_n$ at the different orders in $\mathrm{St}$, which we analyze in detail in what follows.

  \subsubsection*{$\bullet\ O(\mathrm{St}^{-1})\ :$}
   \begin{equation} \label{o0}
    \left(\nabla^2-v^2+d\right)\psi_0=0\ \Longrightarrow\ \psi_0=\ue^{-v^2/2}\xi_0(\bm{x},t)\;,
   \end{equation}
   where $\xi_0$ is still unknown at this stage (we only know it must be normalized to $\pi^{-d/2}=(\int\ud\bm{v}\,\ue^{-v^2})^{-1}$), and will be found by imposing the solvability condition at $O(\mathrm{St}^0)$.

  \subsubsection*{$\bullet\ O(\mathrm{St}^{-1/2})\ :$}
   \begin{eqnarray} \label{o1}
    &\left(\nabla^2-v^2+d\right)\psi_1=\ue^{-v^2/2}v_{\mu}\left[2\sqrt{2}\mathrm{Pe}^{-1/2}\partial_{\mu}-2\sqrt{2}(1-\beta)\mathrm{Pe}^{1/2}u_{\mu}\right]\xi_0\nonumber\\
    &\Longrightarrow\ \psi_1=\ue^{-v^2/2}\left[\xi_1(\bm{x},t)+v_{\mu}\rho_1^{\mu}(\bm{x},t)\right]\nonumber\\
    &\Longrightarrow\ \rho_1^{\mu}=\sqrt{2}(1-\beta)\mathrm{Pe}^{1/2}u_{\mu}\xi_0-\sqrt{2}\mathrm{Pe}^{-1/2}\partial_{\mu}\xi_0\;,
   \end{eqnarray}
   where $\xi_1$ is still unknown at this stage (we only know it must be normalized to $0$), and will be found by imposing the solvability condition at $O(\mathrm{St}^{1/2})$.

  \subsubsection*{$\bullet\ O(\mathrm{St}^0)\ :$}
   \begin{eqnarray} \label{o2}
    \hspace{-2cm}\left(\nabla^2-v^2+d\right)\psi_2&=&\displaystyle-2\frac{L}{TU}\delta(\bm{x})f(\bm{v})\ue^{v^2/2}\nonumber\\
    &&+2\ue^{-v^2/2}\Big\{\left[\ud_t+(2\beta-1)u_{\mu}\partial_{\mu}+(1-\beta)^2\mathrm{Pe}\,u^2\right]\xi_0\nonumber\\
    &&\hspace{1.5cm}+v_{\mu}v_{\nu}\left[2(1-\beta)(\partial_{\mu}u_{\nu})+4(1-\beta)u_{\mu}\partial_{\nu}\right.\nonumber\\
    &&\hspace{2.5cm}\left.-2\mathrm{Pe}^{-1}\partial_{\mu}\partial_{\nu}-2(1-\beta)^2\mathrm{Pe}\,u_{\mu}u_{\nu}\right]\xi_0\nonumber\\
    &&\hspace{1.5cm}\left.+v_{\mu}\left[\sqrt{2}\mathrm{Pe}^{-1/2}\partial_{\mu}-\sqrt{2}(1-\beta)\mathrm{Pe}^{1/2}u_{\mu}\right]\xi_1\right\}
   \end{eqnarray}
   At this point, we have to impose the solvability condition (which would have been trivial at the lower orders (\ref{o0}),(\ref{o1})):
   \begin{eqnarray} \label{csi0}
    &\displaystyle\!\int\!\ud\bm{v}\,\ue^{-v^2/2}\left(\nabla^2-v^2+d\right)\psi_2=\!\int\!\ud\bm{v}\,\psi_2\left(\nabla^2-v^2+d\right)\ue^{-v^2/2}=0\nonumber\\
    &\displaystyle\Longrightarrow\ 0=-2\frac{L}{TU}\delta(\bm{x})\!\int\!\ud\bm{v}\,f(\bm{v})\nonumber\\
    &\displaystyle+2\left[\ud_t+(2\beta-1)u_{\mu}\partial_{\mu}+(1-\beta)^2\mathrm{Pe}\,u^2\right]\xi_0\!\int\!\ud\bm{v}\,\ue^{-v^2}\nonumber\\
    &\displaystyle+2\left[2(1-\beta)(\partial_{\mu}u_{\nu})+4(1-\beta)u_{\mu}\partial_{\nu}-2\mathrm{Pe}^{-1}\partial_{\mu}\partial_{\nu}-2(1-\beta)^2\mathrm{Pe}\,u_{\mu}u_{\nu}\right]\xi_0\!\int\!\ud\bm{v}\,v_{\mu}v_{\nu}\ue^{-v^2}\nonumber\\
    &\displaystyle+2\left[\sqrt{2}\mathrm{Pe}^{-1/2}\partial_{\mu}-\sqrt{2}(1-\beta)\mathrm{Pe}^{1/2}u_{\mu}\right]\xi_1\!\int\!\ud\bm{v}\,v_{\mu}\ue^{-v^2}\nonumber\\
    &\displaystyle\Longrightarrow\ (\ud_t+\bm{u}\cdot\bm{\partial}-\mathrm{Pe}^{-1}\partial^2)\xi_0=\frac{L}{TU}\pi^{-d/2}\delta(\bm{x})\;.
   \end{eqnarray}
   This means that, in the limit $\mathrm{St}\to0$, the nondimensional PDF is
   $$p\to p_0=\ue^{-v^2/2}\psi_0=\ue^{-v^2}\xi_0\;,$$
   where $\xi_0$ behaves as a passive scalar with spatial point source; moreover, after integrating on the covelocity variable to obtain the physical-space density,
   \begin{equation} \label{phsp}
    P(\bm{x},t)\equiv\!\int\!\ud\bm{v}\,p(\bm{x},\bm{v},t)\;,
   \end{equation}
   in dimensional quantities one obtains the well-known equation (\cite{CMAM07}):
   $$(\ud_t+\bm{u}\cdot\bm{\partial}-\kappa\partial^2)P=\frac{1}{T}\delta(\bm{x})\qquad\textrm{(for $\mathrm{St}\to0$)}\;.$$

   Now, substituting $\ud_t\xi_0$ from (\ref{csi0}) back to (\ref{o2}), one gets:
   \begin{eqnarray} \label{p2}
    \left(\nabla^2-v^2+d\right)\psi_2&=&\displaystyle-2\frac{L}{TU}\delta(\bm{x})\left[f(\bm{v})\ue^{v^2/2}-\pi^{-d/2}\ue^{-v^2/2}\right]\nonumber\\
    &&+2\ue^{-v^2/2}\Big\{\left[-2(1-\beta)u_{\mu}\partial_{\mu}+\mathrm{Pe}^{-1}\partial^2+(1-\beta)^2\mathrm{Pe}\,u^2\right]\xi_0\nonumber\\
    &&\hspace{1.5cm}+v_{\mu}v_{\nu}\left[2(1-\beta)(\partial_{\mu}u_{\nu})+4(1-\beta)u_{\mu}\partial_{\nu}\right.\nonumber\\
    &&\hspace{2.5cm}\left.-2\mathrm{Pe}^{-1}\partial_{\mu}\partial_{\nu}-2(1-\beta)^2\mathrm{Pe}\,u_{\mu}u_{\nu}\right]\xi_0\nonumber\\
    &&\hspace{1.5cm}\left.+v_{\mu}\left[\sqrt{2}\mathrm{Pe}^{-1/2}\partial_{\mu}-\sqrt{2}(1-\beta)\mathrm{Pe}^{1/2}u_{\mu}\right]\xi_1\right\}
   \end{eqnarray}
   \begin{equation} \label{psi2}
    \Longrightarrow\ \psi_2=\ue^{-v^2/2}\left[\xi_2(\bm{x},t)+v_{\mu}\rho_2^{\mu}(\bm{x},t)+v_{\mu}v_{\nu}\sigma_2^{\mu\nu}(\bm{x},t)\right]-2\frac{L}{TU}\delta(\bm{x})\phi(\bm{v})
   \end{equation}
   where $\xi_2$, $\rho_2$, $\sigma_2$ can be derived by pushing the investigation at higher orders in $\mathrm{St}$ (not reported here), and $\phi$ satisfies the forced equation
   \begin{equation} \label{fi}
    \left(\nabla^2-v^2+d\right)\phi=f(\bm{v})\ue^{v^2/2}-\pi^{-d/2}\ue^{-v^2/2}\;.
   \end{equation}

  \subsubsection*{$\bullet\ O(\mathrm{St}^{1/2})\ :$}
   \begin{eqnarray} \label{o3}
    \left(\nabla^2-v^2+d\right)\psi_3&=&\displaystyle-2\sqrt{2}\frac{L}{TU}\left[(1-\beta)\mathrm{Pe}^{1/2}u_{\mu}\delta(\bm{x})(\nabla_{\mu}-v_{\mu})\phi(\bm{v})+2\mathrm{Pe}^{-1/2}v_{\mu}\phi(\bm{v})\partial_{\mu}\delta(\bm{x})\right]\nonumber\\
    &&+2\ue^{-v^2/2}\Big\{\left[\ud_t+(2\beta-1)u_{\mu}\partial_{\mu}+(1-\beta)^2\mathrm{Pe}\,u^2\right]\xi_1\nonumber\\
    &&\hspace{1.5cm}+v_{\mu}v_{\nu}\left[2(1-\beta)(\partial_{\mu}u_{\nu})+4(1-\beta)u_{\mu}\partial_{\nu}\right.\nonumber\\
    &&\hspace{2.5cm}\left.-2\mathrm{Pe}^{-1}\partial_{\mu}\partial_{\nu}-2(1-\beta)^2\mathrm{Pe}\,u_{\mu}u_{\nu}\right]\xi_1\nonumber\\
    &&\hspace{1.5cm}+v_{\mu}[(\ldots)\xi_0+(\ldots)\xi_2]\nonumber\\
    &&\hspace{1.5cm}+v_{\mu}v_{\nu}v_{\lambda}(\ldots)\xi_0\Big\}
   \end{eqnarray}
   The solvability condition has to be imposed again at this point:
   \begin{eqnarray} \label{csi1}
    &\displaystyle\!\int\!\ud\bm{v}\,\ue^{-v^2/2}\left(\nabla^2-v^2+d\right)\psi_3=\!\int\!\ud\bm{v}\,\psi_3\left(\nabla^2-v^2+d\right)\ue^{-v^2/2}=0\nonumber\\
    &\displaystyle\Longrightarrow\ 0=-2\sqrt{2}\frac{L}{TU}\!\int\!\ud\bm{v}\,\ue^{-v^2/2}\left[(1-\beta)\mathrm{Pe}^{1/2}u_{\mu}\delta(\bm{x})(\nabla_{\mu}-v_{\mu})\phi(\bm{v})+2\mathrm{Pe}^{-1/2}v_{\mu}\phi(\bm{v})\partial_{\mu}\delta(\bm{x})\right]\nonumber\\
    &\displaystyle+2\left[\ud_t+(2\beta-1)u_{\mu}\partial_{\mu}+(1-\beta)^2\mathrm{Pe}\,u^2\right]\xi_1\!\int\!\ud\bm{v}\,\ue^{-v^2}\nonumber\\
    &\displaystyle+2\left[2(1-\beta)(\partial_{\mu}u_{\nu})+4(1-\beta)u_{\mu}\partial_{\nu}-2\mathrm{Pe}^{-1}\partial_{\mu}\partial_{\nu}-2(1-\beta)^2\mathrm{Pe}\,u_{\mu}u_{\nu}\right]\xi_1\!\int\!\ud\bm{v}\,v_{\mu}v_{\nu}\ue^{-v^2}\nonumber\\
    &\displaystyle+2[(\ldots)\xi_0+(\ldots)\xi_2]\!\int\!\ud\bm{v}\,v_{\mu}\ue^{-v^2}\nonumber\\
    &\displaystyle+2(\ldots)\xi_0\!\int\!\ud\bm{v}\,v_{\mu}v_{\nu}v_{\lambda}\ue^{-v^2}\nonumber\\
    &\displaystyle\Longrightarrow\ (\ud_t+\bm{u}\cdot\bm{\partial}-\mathrm{Pe}^{-1}\partial^2)\xi_1=2\sqrt{2}\frac{L}{TU}\pi^{-d/2}\mathrm{Pe}^{-1/2}[\partial_{\mu}\delta(\bm{x})]\!\int\!\ud\bm{v}\,\ue^{-v^2/2}v_{\mu}\phi(\bm{v})\;.
   \end{eqnarray}
   Let us consider two different cases, corresponding I) to the presence and II) to the absence of rotational symmetry in $f$.\\
   I) In this case $f=f(v)$ that means a centered and isotropic covelocity, corresponding to a source emitting particles with velocity $\beta\bm{u}(\bm{0},t)$ plus isotropic fluctuations (meaningful for $\beta=0$ if $\bm{v}_*=\bm{0}$). In this case, then also $\phi=\phi(v)$ and the integral on the right-hand side (RHS) of (\ref{csi1}) vanishes because of parity, therefore $\xi_1=\textrm{const.}=0$ and (from (\ref{o1}))
   $$\psi_1=\ue^{-v^2/2}v_{\mu}[\sqrt{2}(1-\beta)\mathrm{Pe}^{1/2}u_{\mu}-\sqrt{2}\mathrm{Pe}^{-1/2}\partial_{\mu}]\xi_0\;.$$
   When projecting onto the physical space through (\ref{phsp}), this term integrates to zero, and no explicit correction appears with respect to the tracer approximation:
   $$P=\ue^{-v^2}\xi_0+O(\mathrm{St})\;.$$
   II) On the contrary, if $f(\bm{v})$ has a preferential direction (a mean value $\bm{v}_*$ along the positive $x_d$ coordinate) one has to decompose $\phi=\sum_{l,m}\phi_{(l,m)}Y_{l,m}$ onto the spherical harmonics and obtains a quantum-harmonic-oscillator-like equation. As $f$ shows degeneracy on the azimuthal angle, i.e.\ no preferential direction on the plane orthogonal to the mean emission, the same happens for $\phi$, which thus reduces to a sum of Legendre polynomials (spherical harmonics with $m=0$). With the substitution into the RHS of (\ref{csi1}) in mind, one can discard the isotropic sector and, if e.g.
   $$f(\bm{v})=(2\pi\sigma^2)^{-d/2}\ue^{-(\bm{v}-\bm{v}_*)^2/2\sigma^2}\;,$$
   one gets:
   \begin{eqnarray*}
    &[\partial^2_v+(d-1)v^{-1}\partial_v-l(l+d-2)v^{-2}-v^2+d]\phi_{(l)}(v)=\ue^{v^2/2}f_{(l)}(v)=\\
    &=\ui^l\sqrt{4\pi(2l+1)}(2\pi\sigma^2)^{-d/2}\ue^{-(1-\sigma^2)v^2/2\sigma^2-v_*^2/2\sigma^2}j_l(-2\ui vv_*/\sigma^2)
   \end{eqnarray*}
   (with $l\ge1$), $j_l$ being the spherical Bessel function of the first kind. Thus,
   \begin{eqnarray} \label{fil}
    \phi_{(l)}(v)&=&\displaystyle v^l\ue^{-v^2/2}\left\{\mathcal{U}\left(\frac{l}{2},\frac{d}{2}+l,v^2\right)\left[C_{(l)}+K\!\int_0^v\ud k\,k^{l+d-1}\mathcal{L}_{-l/2}^{(d/2+l-1)}(k^2)f_{(l)}(k)\right]\right.\nonumber\\
    &&\displaystyle\left.+\mathcal{L}_{-l/2}^{(d/2+l-1)}(v^2)\left[c_{(l)}-K\!\int_{\infty}^v\ud k\,k^{l+d-1}\mathcal{U}\left(\frac{l}{2},\frac{d}{2}+l,k^2\right)f_{(l)}(k)\right]\right\}\;,
   \end{eqnarray}
   where the values of $C_{(l)}$ and $c_{(l)}$ can be found by imposing regularity and normalization, and the constant $K$ is given by:
   $$K\equiv\frac{k^{-d-2l}\ue^{k^2}}{\displaystyle 2\mathcal{U}\left(\frac{l}{2},\frac{d}{2}+l,k^2\right)\mathcal{L}_{-l/2-1}^{(d/2+l)}(k^2)-l\mathcal{U}\left(1+\frac{l}{2},1+\frac{d}{2}+l,k^2\right)\mathcal{L}_{-l/2}^{(d/2+l-1)}(k^2)}$$
   (whose value for $d=3$ and $l=1$ is $\pi/2$). Here, $\mathcal{L}$ is the generalized Laguerre function and $\mathcal{U}$ is Kummer's confluent hypergeometric function (of the second kind) (\cite{GR65}).\\
   One can further focus on the first anisotropic sector $l=1$ (the only non-vanishing when substituting in (\ref{csi1})), which implies $C_{(1)}=0=c_{(1)}$ for regularity.%
   \footnote{Actually, $\phi$ itself needs not be regular in the $\bm{v}$ coordinate, but the condition on the integration constants $C_{(1)}$ and $c_{(1)}$ follows from plugging (\ref{fil}) into (\ref{psi2}) and (\ref{herm}), and from the consideration that $p_2$ must be regular for $v\to0$ and vanishing for $v\to\infty$.}
   As a result, the RHS of (\ref{csi1}) reduces to
   \begin{equation} \label{simpl}
    4\sqrt{\frac{2}{3}}\frac{L}{TU}\pi^{-(d-1)/2}\mathrm{Pe}^{-1/2}[\partial_{x_d}\delta(\bm{x})]\!\int\!\ud v\,v^d\ue^{-v^2/2}\phi_{(1)}(v)\;,
   \end{equation}
   the latter integral being just a number.

 \section{Conclusions} \label{sec:conc}

 Gathering all of our information, for the particle PDF we have:
 \begin{eqnarray} \label{resume}
  p(\bm{x},\bm{v},t)&=&\ue^{-v^2/2}\left[\psi_0(\bm{x},\bm{v},t)+\mathrm{St}^{1/2}\psi_1(\bm{x},\bm{v},t)\right]+O(\mathrm{St})\\
  &=&\ue^{-v^2}\xi_0(\bm{x},t)+\mathrm{St}^{1/2}\ue^{-v^2}\Big\{\xi_1(\bm{x},t)+v_{\mu}\times\nonumber\\
  &&\left.\times\left[\sqrt{2}(1-\beta)\mathrm{Pe}^{1/2}u_{\mu}(\bm{x},t)\xi_0(\bm{x},t)-\sqrt{2}\mathrm{Pe}^{-1/2}\partial_{\mu}\xi_0(\bm{x},t)\right]\right\}+O(\mathrm{St})\;,\nonumber
 \end{eqnarray}
 with $\xi_0$ and $\xi_1$ obeying the forced advection--diffusion equations (\ref{csi0}) and (\ref{csi1}) (simplified through (\ref{simpl})) respectively.\\
 From (\ref{resume}), one can obtain information on different physical properties, or otherwise deduce the physical-space PDF (\ref{phsp}) directly:
 \begin{equation} \label{spfi}
  P(\bm{x},t)=\pi^{d/2}\xi_0(\bm{x},t)+\mathrm{St}^{1/2}\pi^{d/2}\xi_1(\bm{x},t)+O(\mathrm{St})\;.
 \end{equation}

 Notice that, up to the order we have investigated ($O(\mathrm{St}^{1/2})$), gravity plays no role. In other words, particle sedimentation, which is known to take place with a (bare) terminal velocity $\propto\mathrm{St}$ plus flow-induced corrections, thus gives rise to a subleading effect.

 In order to give a practical example of application of our equations, let us concentrate on micro-powder dispersion in microchannels. In a typical microchannel size of $L\sim10^{-4}\,\mathrm{m}$, inside which distilled water at temperature $T\sim320\,\mathrm{K}$ (density $\rho_{\mathrm{f}}\sim988\,\mathrm{kg}\,\mathrm{m}^{-3}$ and kinematic viscosity $\nu\sim5\;10^{-7}\,\mathrm{m}^2\,\mathrm{s}^{-1}$) is flowing at the typical velocity $U\sim10^{-4}\,\mathrm{m}\,\mathrm{s}^{-1}$, we inject lead micro-powder containing particles of size $R\sim2\;10^{-6}\,\mathrm{m}$ having a density $\rho_{\mathrm{p}}\sim11.3\;10^3\,\mathrm{kg}\,\mathrm{m}^{-3}$. The Stokes time is
 $\tau\sim2\;10^{-5}\,\mathrm{s}$ and the particle diffusivity is $\kappa\simeq k_{\mathrm{B}}T/(6\pi\nu\rho_{\mathrm{f}}R)\sim2.4\;10^{-13}\,\mathrm{m}^2\,\mathrm{s}^{-1}$ (from Einstein's relation, $k_{\mathrm{B}}=1.38\;10^{-23}\,\mathrm{J}\,\mathrm{K^{-1}}$ being Boltzmann's constant). In this way, the P\'eclet number is $\mathrm{Pe}\equiv LU/\kappa\sim4\;10^4$, the Stokes number is $\mathrm{St}\equiv\tau U/L\sim2\;10^{-5}$, and thus $\mathrm{St}\,\mathrm{Pe}=O(1)$, according to (\ref{loosely}). Note that, despite the fact that $\mathrm{St}$ is quite small, the first correction in our equation does appear at $O(St^{1/2})$, thus giving a more appreciable $5\;10^{-3}$ which perfectly falls in the present perturbative scheme.

 \begin{acknowledgments}
  We thank A.\ Celani, P.\ Muratore-Ginanneschi and D.\ Vincenzi for useful discussions and suggestions.
 \end{acknowledgments}

 \oneappendix

 \section{The Fokker--Planck equation associated to the Lagrangian description}

 The derivation of the Fokker--Planck equation (\ref{fp}) from the Lagrangian equations (\ref{dyn}) is briefly recalled here by exploiting the corresponding unforced (sourceless) case.\\
 The distribution $p(\bm{x},\bm{v},t)$ is given by the average (on the realizations of the random noise $\bm{\eta}$) of the product of two Dirac deltas, expressing the probability density that, at time $t$, the particle location $\bm{X}$ and covelocity $\bm{V}$ equal the spatial coordinate $\bm{x}$ and the covelocity variable $\bm{v}$, respectively:
 \begin{equation} \label{def}
  p(\bm{x},\bm{v},t)\equiv\langle\delta(\bm{x}-\bm{X}(t))\delta(\bm{v}-\bm{V}(t))\rangle
 \end{equation}
 (notice that, in this expression and in the following ones, the average acts on the random functions $\bm{X}$ and $\bm{V}$, which are $\bm{\eta}$-dependent, and not on $\bm{x}$ and $\bm{v}$). A differentiation in time and the use of the chain rule give:
 $$\frac{\partial p}{\partial t}=\left\langle\dot{\bm{X}}(t)\cdot\frac{\partial\delta(\bm{x}-\bm{X}(t))}{\partial\bm{X}(t)}\delta(\bm{v}-\bm{V}(t))+\delta(\bm{x}-\bm{X}(t))\dot{\bm{V}}(t)\cdot\frac{\partial\delta(\bm{v}-\bm{V}(t))}{\partial\bm{V}(t)}\right\rangle\;.$$
 Exploiting (\ref{dyn}) and of the translational invariance of $\delta$, but keeping the term explicitly depending on $\bm{\eta}$ momentarily apart, we get:
 \begin{eqnarray*}
  \frac{\partial p}{\partial t}&=&\Bigg\langle\{\bm{V}(t)+\beta\bm{u}(\bm{X}(t),t)\}\cdot\left[-\frac{\partial\delta(\bm{x}-\bm{X}(t))}{\partial\bm{x}}\right]\delta(\bm{v}-\bm{V}(t))+\delta(\bm{x}-\bm{X}(t))\times\\
  &&\left.\times\left\{-\frac{\bm{V}-(1-\beta)\bm{u}[\bm{X}(t),t]}{\tau}+(1-\beta)\bm{g}\right\}\cdot\left[-\frac{\partial\delta(\bm{v}-\bm{V}(t))}{\partial\bm{v}}\right]\right\rangle+\langle\bm{\eta}(t)\ldots\rangle\;.
 \end{eqnarray*}
 The derivatives can now be moved out of the brackets and, inside the curly braces, the capital letters can be replaced by the corresponding small ones because of the deltas:
 \begin{eqnarray*}
  \frac{\partial p}{\partial t}&=&-\frac{\partial}{\partial x_{\mu}}\left\langle\{v_{\mu}+\beta u_{\mu}(\bm{x},t)\}\delta(\bm{x}-\bm{X}(t))\delta(\bm{v}-\bm{V}(t))\right\rangle\\
  &&-\frac{\partial}{\partial v_{\mu}}\left\langle\left\{-\frac{v_{\mu}-(1-\beta)u_{\mu}(\bm{x},t)}{\tau}+(1-\beta)g_{\mu}\right\}\delta(\bm{x}-\bm{X}(t))\delta(\bm{v}-\bm{V}(t))\right\rangle\\
  &&+\langle\bm{\eta}(t)\ldots\rangle\;.
 \end{eqnarray*}
 Moving the curly brackets out of the averages, applying (\ref{def}), and moving to the left-hand side (LHS) the first two terms on the RHS, we are left with:
 \begin{eqnarray} \label{int}
  &\displaystyle\left[\frac{\partial}{\partial t}+\frac{\partial}{\partial x_{\mu}}(v_{\mu}+\beta u_{\mu}(\bm{x},t))+\frac{\partial}{\partial v_{\mu}}\left(\frac{(1-\beta)u_{\mu}(\bm{x},t)-v_{\mu}}{\tau}+(1-\beta)g_{\mu}\right)\right]p=\nonumber\\
  &=\displaystyle\left\langle\delta(\bm{x}-\bm{X}(t))\frac{\sqrt{2\kappa}}{\tau}\eta_{\mu}(t)\frac{\partial\delta(\bm{v}-\bm{V}(t))}{\partial V_{\mu}(t)}\right\rangle\equiv[*]\;.
 \end{eqnarray}
 If one takes into account the source contribution, which forces the addition of a term $\delta(\bm{x})f(\bm{v})/T$ on the RHS of (\ref{int}) (this can be interpreted as the imposition of a boundary condition), the derivation of (\ref{fp}) is thus complete once the stochastic term $[*]$ is made explicit, which can easily be done by means of a Gaussian integration by parts (here, $\mathrm{D}$ represents the functional derivative) (\cite{F95}):
 \begin{eqnarray*}
  [*]&=&\frac{\sqrt{2\kappa}}{\tau}\!\int_{-\infty}^{\infty}\ud t'\,\langle\eta_{\mu}(t)\eta_{\nu}(t')\rangle\left\langle\frac{\mathrm{D}}{\mathrm{D}\eta_{\nu}(t')}\left[-\delta(\bm{x}-\bm{X}(t))\frac{\partial\delta(\bm{v}-\bm{V}(t))}{\partial v_{\mu}}\right]\right\rangle\\
  &=&-\frac{\sqrt{2\kappa}}{\tau}\frac{\partial}{\partial v_{\mu}}\!\int_{-\infty}^{\infty}\ud t'\,\delta_{\mu\nu}\delta(t-t')\left\langle\frac{\mathrm{D}[\delta(\bm{x}-\bm{X}(t))\delta(\bm{v}-\bm{V}(t))]}{\mathrm{D}\eta_{\nu}(t')}\right\rangle\\
  &=&-\frac{\sqrt{2\kappa}}{\tau}\frac{\partial}{\partial v_{\mu}}\left\langle\frac{\partial\delta(\bm{x}-\bm{X}(t))}{\partial X_{\lambda}(t)}\frac{\mathrm{D}X_{\lambda}(t)}{\mathrm{D}\eta_{\mu}(t)}\delta(\bm{v}-\bm{V}(t))+\delta(\bm{x}-\bm{X}(t))\frac{\partial\delta(\bm{v}-\bm{V}(t))}{\partial V_{\lambda}(t)}\frac{\mathrm{D}V_{\lambda}(t)}{\mathrm{D}\eta_{\mu}(t)}\right\rangle\;.
 \end{eqnarray*}
 Integrating (\ref{dyn}) formally in time and invoking causality (a variation of $\bm{\eta}$ at time $t$ cannot cause any change in $\bm{X}$ or $\bm{V}$ at the same instant, but only at later times), in the last expression the former functional derivative vanishes and, in the latter, only the term explicitly showing $\bm{\eta}$ survives:
 \begin{eqnarray*}
  [*]&=&-\frac{\sqrt{2\kappa}}{\tau}\frac{\partial}{\partial v_{\mu}}\left\langle-\delta(\bm{x}-\bm{X}(t))\frac{\partial\delta(\bm{v}-\bm{V}(t))}{\partial v_{\lambda}}\!\int_{-\infty}^t\ud t''\frac{\sqrt{2\kappa}}{\tau}\frac{\mathrm{D}\eta_{\lambda}(t'')}{\mathrm{D}\eta_{\mu}(t)}\right\rangle\\
  &=&\frac{2\kappa}{\tau^2}\frac{\partial^2}{\partial v_{\mu}\partial v_{\lambda}}\langle\delta(\bm{x}-\bm{X}(t))\delta(\bm{v}-\bm{V}(t))\rangle\!\int_{-\infty}^{\infty}\ud t''\theta(t-t'')\delta_{\mu\lambda}\delta(t-t'')\\
  &=&\frac{\kappa}{\tau^2}\frac{\partial^2}{\partial v_{\mu}\partial v_{\mu}}p
 \end{eqnarray*}
 (the Heaviside theta is indeed such that $\theta(0)=1/2$).

\end{document}